\documentclass[conference]{IEEEtran}
\IEEEoverridecommandlockouts
\usepackage{cite}
\usepackage{amsmath,amssymb,amsfonts}
\usepackage{algorithmic}
\usepackage{graphicx}
\usepackage{textcomp}
\usepackage{xcolor}
\usepackage{subcaption}
\usepackage{url}
\def\BibTeX{{\rm B\kern-.05em{\sc i\kern-.025em b}\kern-.08em
    T\kern-.1667em\lower.7ex\hbox{E}\kern-.125emX}}
\begin{document}

\title{Autonomous Low Power IoT System Architecture for Cybersecurity Monitoring}

\author{\IEEEauthorblockN{Zag ElSayed$^{\star}$, Nelly Elsayed$^{\dagger}$, Chengcheng Li$^{\dagger}$, Magdy Bayoumi$^{\ddagger}$}
\IEEEauthorblockA{\textit{$^{\star \dagger}$School of Information Technology} \\
\textit{$^{\ddagger}$Department of Electrical and Computer Engineering}\\
\textit{$^{\star}$Oil Center Research, llc, USA} \\
\textit{$^{\star \dagger}$University of Cincinnati, OH, USA}\\
\textit{$^{\ddagger}$University of Louisiana at Lafayette, LA, USA}\\
}

}
\thispagestyle{empty}

\begin{huge}
	IEEE Copyright Notice
\end{huge}

\vspace{5mm} 

\begin{large}
	Copyright (c) 2022 IEEE
\end{large}

\vspace{5mm} 

\begin{large}
	Personal use of this material is permitted. Permission from IEEE must be obtained for all other uses, in any current or future media, including reprinting/republishing this material for advertising or promotional purposes, creating new collective works, for resale or redistribution to servers or lists, or reuse of any copyrighted component of this work in other works.
\end{large}

\vspace{5mm} 

\begin{large}
	\textbf{Accepted to be published in:} IEEE WFIoT-2022; 26 October - 11 November , 2022 - Yokohoma, Japan.
	https://wfiot2022.iot.ieee.org/
	
\end{large}

\vspace{5mm} 

\maketitle

\begin{abstract}
Network security morning (NSM) is essential for any cybersecurity system, where the average cost of a cyber attack is \$1.1 million. No matter how secure a system, it will eventually fail without proper and continuous monitoring. No wonder that the cybersecurity market is expected to grow up to \$170.4 billion in 2022. However, the majority of legacy industries do not invest in NSM implementation until it is too late due to the initial and operation costs and static unutilized resources. Thus, this paper proposes a novel dynamic Internet of things (IoT) architecture for an industrial NSM that features a low installation and operation cost, low power consumption, intelligent organization behavior, and environmentally friendly operation. As a case study, the system is implemented in a mid-range oil a gas manufacturing facility in the southern states with more than 300 machines and servers over three remote locations and a production plant that features a challenging atmosphere condition. The proposed system successfully shows a significant saving ($>$65\%) in power consumption, acquires one-tenth of the installation cost, develops an intelligent operation expert system tool as well as saves the environment from more than 500mg of CO2 pollution per hour, promoting green IoT systems. 

\end{abstract}


\section{Introduction}

Network Security Monitoring (NSM) is defined as the collection, detection, and analysis of network security data as well as escalation of indications and warnings to detect and respond to intrusions on computer networks~\cite{1}. Network security monitoring tools typically feature: network-based threat detection, machine base threat detection, proactive network queries for security data and “hunting” for suspicious behavior, integration with one or more threat feeds, and create security alerts~\cite{2}. Information Network Security has traditionally started via the United States Department of Defense (US DoD) categorizes the domains of Computer Network Defense (CND)~\cite{3}.

NSM is based upon the concept that prevention eventually fails. No matter how much time and resources were invested in static securing a network, without employing a continuous monitoring operation, eventually, the scenario will make the bad guys win. By analogy, all Middle Ages castles eventually were fallen or surrendered due to advanced weapon technology or political events~\cite{bejtlich2013practice}. Thus, when this happens, there should be an organized technical system able to detect and respond to the intruder’s presence so that an incident may be declared and the intruder can be eradicated with minimal damage done.

\begin{figure}
	\centerline{\includegraphics[width=8.7cm, height= 7 cm]{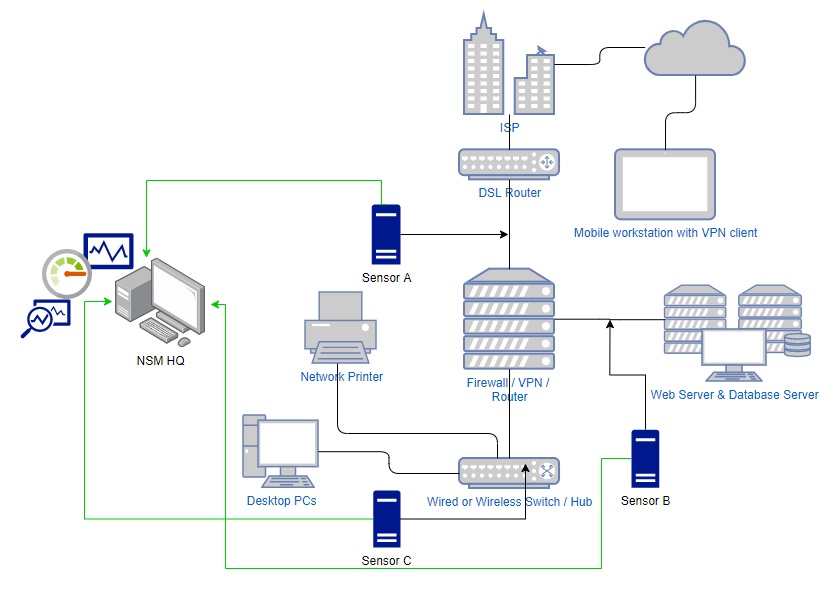}}
	\caption{A typical NSM sensor placement scenario.}
	\label{figure1}
\end{figure}

Any NSM system essentially depends on a device that captures network traffic, detects the anomaly, and performs analysis over various levels of detail. This device is called an NSM sensor. NSM sensors consist of a software suite that is very resource hungry and relay on expensive hardware. The storage disk is the main issue with the sensor, which can grow up to $\sim$1Tbyte per day in some situations depending on the data type that it uses. Additionally, NSM data can grow exponentially and require regularly scheduled maintenance, backup, and means of accessibility. It is important to note that if the captured data is lost, this can limit the ability to perform retrospective analysis, which is crucial for a current investigation. Most of the current technologies consider the sensors as passive devices with two interfaces, one interface for management and logging and the other interface for traffic capture. The sensor is also usually used for just reporting to a centralized point for analysis and alert reporting, such as the Snort$\copyright$ repository, as shown in Fig.~\ref{figure1}, where the green lines represent the management traffic to the central NSM HQ. The static operation of the sensors makes them frozen in time, role, and functionality, as well as the lack of feedback control, which makes the sensor look like a waste of investment that pushes back many industrial implementations to refuse to add dedicated NSM systems to their network until it is too late, which is one of the main issues with the NSM implementation. 

However, by boosting "life" to these sensors via communication, collaboration, and active control of sensor system architecture can increase their efficiency, increase their illusion of intelligence and reduce the overhead of operation and maintenance costs. Thus, this work applies a novel approach by injecting the dynamic Internet of Things (IoT) concepts to the NSM sensors, which reduces their size, adds the communication and the control framework, and applies the messaging system to reduce the hardware requirements, lower the operating power consumption and makes the detection and prevention faster for many network intrusions.

Additionally, as a proof of concept, this architecture was applied to the information system network of an oil and gas production facility with more than 300 machines and serves, serving three remote branches. The proposed architecture system saves more than one order of magnitude in equipment cost and more than 1.867MW of annual power consumption, as well as saves the environment from more than 4000 mg of CO2 emission per day.

\section{Background}
Oil and gas industrials support 10.3 million jobs in the United States and nearly eight percent of our nation's gross domestic product, with 32.5\% of the market share. The oil and gas industry faces unique cybersecurity challenges, given their distributed, decentralized structures and the large operational technology environment that does not fit the traditional cybersecurity scenarios. Thus, the majority of the manufacturers do not have a complete cyber security implementation due to cost, revenue, utilization, and investment. The investment gap has left most heavy industries insufficiently prepared for monitoring, detecting, and preventing threats. As a result, they are attractively targeted by cybercrimes; in 2018, nearly 60\% of relevant surveyed organizations had experienced a breach that ended up with financial loss, several of which considered adding the NSM system only after the cyberattack incidents~\cite{4}.

\begin{figure}
	\centerline{\includegraphics[width=8.7cm, height= 11 cm]{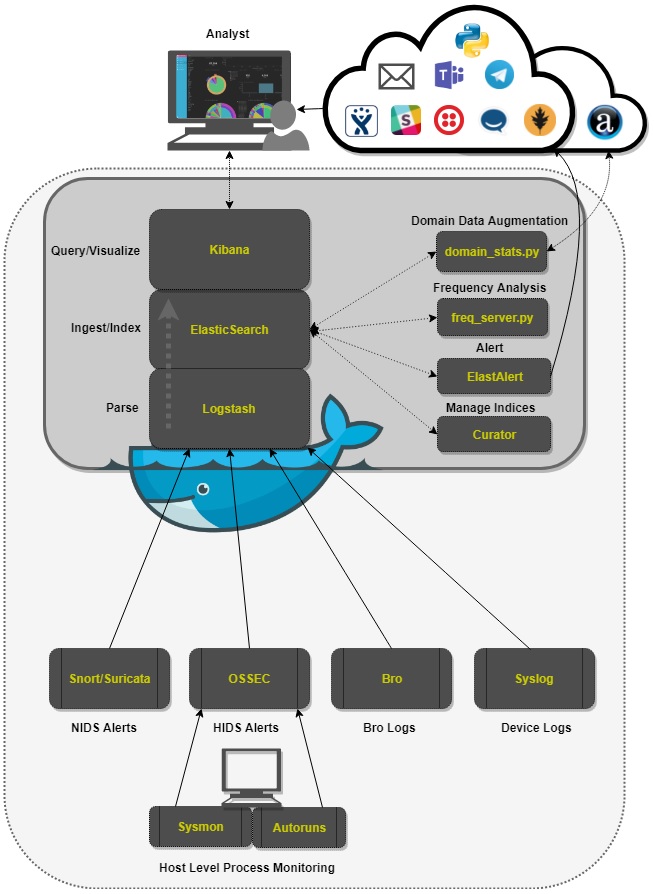}}
	\caption{Security Onion high-level system architecture~\cite{9}.}
	\label{figure2}
\end{figure}

The current technology features many advanced and sophisticated NSW systems for open source and commercial implementations~\cite{5}. However, no matter how complex the NSM system is, it still depends on the essential actor of the system, which is the NSM Sensor Platform (SP). The SP is a combination of hardware and software that perform: collection such as Packet Capturing (PCAP) or NetFlow~\cite{6}, detection such as Signature-Based, Reputation-Based, and/or Anomaly-Based, and network threat analysis~\cite{7}. SPs can be classified by their functionality into three classes: Collection Only, Half-Cycle, or Full-Cycle, depending on what operations they perform, which can be: collection only, collection and detection, or collection, detection, and analysis, respectively.

SPs usually require a lot of hardware resources. For example, a simple Security Onion$\copyright$ SP requires 12GB of memory, four cores processor, 200GB of disk storage, and two network interfaces~\cite{8}. Security onion system architecture is shown in Fig.~\ref{figure2}. Security Onion is a free and open-source Linux distribution for threat hunting, enterprise security monitoring, and log management. The Security Onion includes Elasticsearch~\cite{elasticsearch2018elasticsearch}, Logstash, Kibana, Suricata, Zeek, Wazuh, Stenographer, Hive, Cortex, CyberChef, and NetworkMiner, and it requires expensive hardware configuration~\cite{9}.

\begin{figure}
	\centerline{\includegraphics[width=8.7cm, height= 5 cm]{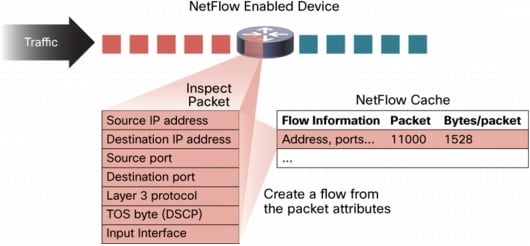}}
	\caption{NetFlow cash data sample with network behavior information~\cite{9}.}
	\label{figure3}
\end{figure}

NetFlow is an embedded instrumentation within Cisco IOS Software. It is used to characterize network operation and vision into the network, which is an essential tool for IT and system analysts [10]. In response to new requirements and pressures, network operators are finding it critical to understand how the network is behaving, including application, network usage, network productivity, anomaly, and security vulnerabilities. A sample of the NetFlow data structure record is shown in Fig.~\ref{figure3}.

NetFlow protocol is very useful; however, it consumes a high bandwidth on the network, it is vendor specific, version specific, and it also increases the network devices processor utilization (by around $\sim$20\%)~\cite{10} while reducing the cache availability that is highly depending on the network performance, especially during peak hours which in the case study is between 8:00 am to 9:30 am, 1:30 pm to 2:30 pm as well as on major social and/or political events (e.g., elections).
SPs are added to the network via two primary methods. The first method is done via port mirroring, which requires some reconfiguring of the network device such as switches and routers (which is not very suitable for the majority of established industrial implementations) that many system network administrators would refuse to perform without proper testing and several simulation runs of the whole network computer system, especially for the production automation facilities where a heterogeneous network profile of different sensor, actuators gauges, and automation devices are most probably installed and configured with less than minimum documentation and little options activated. The second method is via a network tapping that can be more transparent to the network administration and management team, where network taps could be implemented via a hardware tap or virtually via a software tap~\cite{11,12}. Software tap is usually preferred for temporary solutions and remote installations. Both types of network taps provide basic access to the wired network lines to capture the outbound (Tx) and/or the inbound (Rx) data traffic. The data are basically seen as packets from the NSM level of operation in the TCP/IP stack. The essential data types the SPs process are:

\begin{itemize}
	\item \textit{Full Packet Capture in from of (PCAP):} has all the details of the communication between two nodes.
	\item \textit{Flow Data (FD):} has the log of the communication and the summary log of the sessions.
	\item \textit{Packet String Data (PSTR):} has the human-readable contents of the packets.
\end{itemize}

The associate relative data size of the resulted log is usually represented as 100\%, 0.01\%, and 4\%, respectively, where the PCAP is 100\%. Their ratios also reflect and determine the required storage size and the storage management protocols.

The Internet of Things refers to the ever-growing network of physical objects that feature identifiers for internet connectivity and data exchange communication that occurs between these objects and other systems [13]. This work proposes a novel NSM architecture based on the IoT concept that converts the static NSM sensors into active IoT sensors framework, applying the concept of the NSM Hub and the IoT clouds. The proposed sensors are built on miniature board machines. The IoT hub is implemented on a single NSM machine with a backup (to avoid a single point of failure) and an IoT cloud storage. The proposed system overcomes the cost associated with traditional NSM via reduced hardware, increased sensor utilization, and saved power via low operational energy consumption.

\begin{figure}
	\centerline{\includegraphics[width=8.7cm, height= 8 cm]{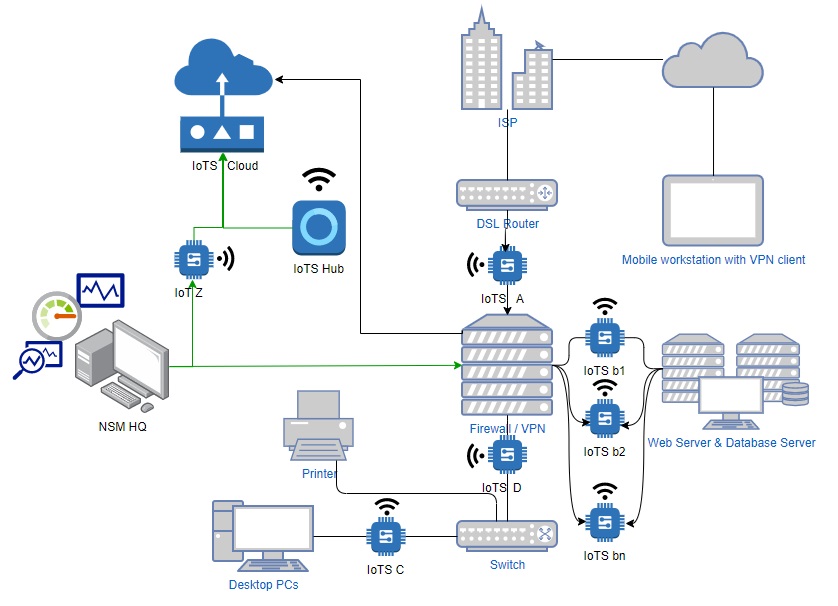}}
	\caption{The proposed system IoTS NSM highlevel architeture.}
	\label{figure4}
\end{figure}

\section{System Design and Silence Unveil}
\subsection{Proposed Architecture}
The higher level of the proposed system architecture is shown in Fig.~\ref{figure4} where it shows the sensors were replaced by IoT Network Monitoring Sensor (IoTS) sensors, an IoT Hub was added for data aggregation, and an IoT Could storage was established. The following sub-sections discuss the detailed descriptions, excluding cloud architecture as it is a standard implementation and out of the scope of this work.

\begin{figure}
	\centerline{\includegraphics[width=8.7cm, height= 4 cm]{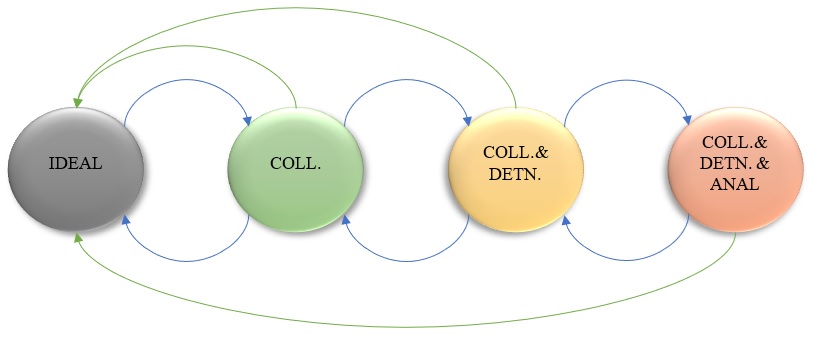}}
	\caption{IoTS node action roll state machine with power saving signal transmission via green arrows and state analysis escalation signals via blue arrows.}
	\label{figure5}
\end{figure}

\subsection{The IoTS NSM Sensor}

This work proposes a miniature compact dynamic NSM sensor (i.e., IoTS) to replace the traditional static machines (servers that are NSM Sensors). The Proposed IoTS hardware is a custom Raspberry Pi build with two-gigabit ethernet ports and a Wi-Fi Antenna. The network ports are used for software tapping and capturing packets. The IoTS uses Wi-Fi as a primary link for neighbor discovery and IoT Hub control signals. In the case study, the IoTS node is selected with the specs as shown in Table~\ref{Table1}. Additionally, it is crucial to notice that with this hardware configuration, the node cost does not exceed one-tenth the cost of a traditional midrange NSM Sensor.  

 IoTS uses a compiled NSM software for packet capture, where the session packet capture process is done via FProbe, the PCAP is performed via daemonlogger, and the PSTR is based on URLsnarf software~\cite{14}. The IoTS node features the sensor functionality polymorphism. It is characterized by the ability to change the type of the packet, capturing the scope and the sensor role of the node according to the IoT Hub control signal that acts as a transition trigger. The IoTS role can switch between collection only, collection and detection, or collection and detection and analysis, as shown in Fig.~\ref{figure5}, role state machine. 
 
 \begin{table}
 
 	\begin{tabular}{|l|l|}
 		\hline
 	\textbf{Parameter}	& \textbf{Hardware Node Specs}  \\
 	\hline
CPU	&Broadcom quad-core 64-bit SoC, 1.5GHz\\
\hline
Memory&	8GB LPDDR4 \\
\hline
NICs&	2x  Gigabit Ethernet, Endace\\
\hline
Wireless&	5.0 GHz IEEE 802.11b/g/n/ac, Bluetooth 5.0, BLE\\
\hline
Storage	&2TB USB 3.0 Flash Drive, w/r 100MB/s\\
\hline
Pwr Consum&	Idea 1Watt, Peak 1.5Watt, Fanless\\
\hline

 	\end{tabular}
 \caption{IoTS Node Hardware Specs.}
 	\label{Table1}
 \end{table}

\subsection{The IoT Hub Sensor}
To all appearances, the IoT Hub could be seen as a regular NSM sensor machine from the hardware configuration point. However, the proposed implementation adds some IT intelligence that manages the IoTS states based on the detection warning and alerts, saves the power state for the least needed operation, and directs the data storage location when needed via the neighbor storage sharing mechanism (NSSM) which uses Dijkstra shortest pass algorithm the find nearby nodes (using network hops and network utilization as a metric) to find the nearest available node with extra storage to maneuver the data storage task when needed, schedules the data transfer to the IoT cloud, adds a second level detection alerts filtering to reduce false positive alerts via a pooling mechanism to increase the system's precision.

The IoT Hub also allows NSM HQ connection for flow monitoring and operation to an individual IoTS node or any other part of the system, including a communication route scenario. Additionally, the proposed IoTS enriches the monitoring capability to the point that even the management flow could be monitored for an extra security precaution with very low installation and operation cost with a low network configuration overhead. The IoT Hub uses the following criteria for heuristic information and decision-making mechanism to manage the state machine of each node in the network domain based on the data that it gathers from the node messaging system, such as sensor state, sensor role state, anomaly traffic detection, attack attempt detection, nearby nodes graph, and data collection location change. These data are stored in a database on the cloud with an instance cached on the IoT Hub for fast access. The database contains weight and probability values that are structured into a string that describes the whole system as a unit and assigns a proposed list of actions to be deployed to the IoTS to either change or keep their role state as well as predict the NSSM triggers. 

The initial values of the system states are set manually to the IoT Hub. The IoT state is initially set to coll. \& detn, (i.e., allowing the global Collection and Detection) for all the nodes. However, it could be configured according to the system operation needs, such as normal routine monitoring, warning threshold checks, anomaly detection, or attack attempt occurrence. Additionally, the system representation string database is used by a parametrized generic algorithm~\cite{15,16} that provides a simple and primitive form of advice assistance to the network monitoring analysts team that is based on a supervised learning scheme that gains experience with time and can provide a smart performance on the long run. This unit has added the system as a first step toward a more tailored and case-specific informed decision-making mechanism for a fully automated IoTS system.

The proposed system was implemented at an oil and gas mid-range production plant with 15 IoTS nodes distributed at the production plant, the sales and vendor access locations, the server room as well as three remote branches in Louisiana, Texas, and Oklahoma. The system was observed for $\sim$11 months and during an extremely hot summer ($>$100$^{\circ}$F) under dusty and greasy operation conditions.

\begin{figure}
	\centerline{\includegraphics[width=8.7cm, height= 5 cm]{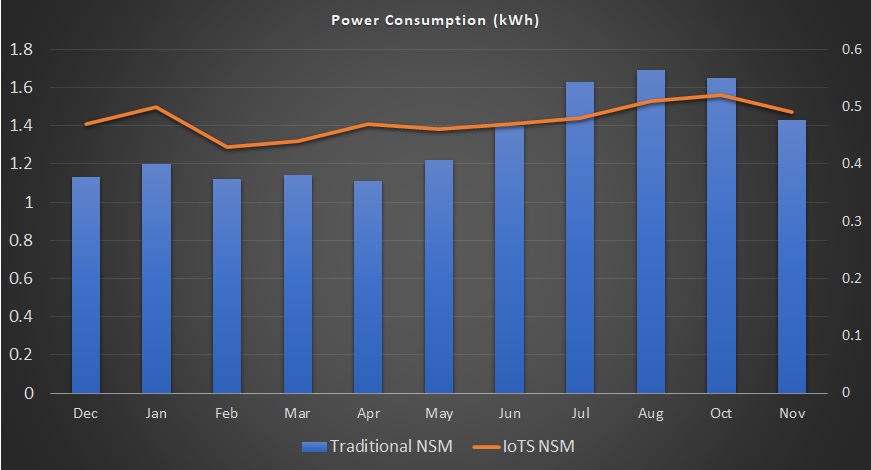}}
	\caption{Power consumption comparison between traditional NSM system and a substituted IoTS NSM system during eleven months test period.}
	\label{figure6}
\end{figure}

\section{Results and Observation}
The IoT system concept enhances the static NSM architect with dynamic behavior. It helps the network monitoring analysts by adding more diverse operations via new functionalities such as dynamic sensor role change, power saving options, and sensor agents-like smart behavior. The major enhancements, as seen from the industrial management point of view, are lower power consumption and reduced initialization, deployment, and operation cost, which are the main barriers that delay the NSM implementation for this sector of the industry, as well as the harsh environment operation condition tolerance.

The initial cost was reduced by order of magnitude. The low-power and fan-less hardware boards are the winning solutions for the outdoors and unbearable grease and dust particles contamination the atmosphere. During 11 month test period, the power consumption was recorded and is shown in Fig.~\ref{figure6}, where the blue bars indicate the traditional NSM power consumption on the left scale vertical axis and the yellow line indicates the IoTS NSM power consumption on the right vertical scale axis. Additionally, the average power computation during the eleven months operation was reduced from 1.38kWh to 0.48kWh (by $\sim$65\% less) at a peak sensor utilization that led to saving the environment from 563.4 mg of CO2 per hour, making the proposed system a green IoT solution.

\section{Conclusion}
The IoTS NSM system architecture approach demonstrates its efficiency and promotes an environmentally friendly solution, especially in the major polluting industries where its middle-range sector is facing many challenges in entering the next cyber information age era. The proposed system characterizes by the IoT architecture capabilities that give the network security analysts a new boost and novel tools of network security monitoring operation, optimization and functionality, cooperative threat detection and prevention mechanisms with a minimum effort from the system administration teams, and zero configuration tasks to the network engineers. 

However, the proposed system has several limitations. First, regarding the physical security of the nodes, which is the case with all wireless sensor networks. Second, the proposed system may suffer from packet drop that occurs during the transition time of an IoTS during the role transfer operation, which can take up to 35 seconds in the worst case, which could be solved by deploying a multiple IoTSs on the same flow line to back up the transfer. Second, the quality of the hardware plays a crucial role, especially the NICs, in preventing buffer overflow and packet loss which may jeopardize the precision of the NSM detection system. Finally, the parametrized learning algorithm needs high attention from the system analyst team, and such learning methods are pruned to learn wrong decisions, which can be improved via adding a more sophisticated but lightweight learning technology such as adaptive pattern recognition such as Artificial Immune System, which the authors are considering for the future work.

\bibliographystyle{ieeetr}
\bibliography{CybersecurityMonitoringReferences}

\begin{thebibliography}{10}

\bibitem{1}
C.~Sanders and J.~Smith, {\em Applied network security monitoring: collection,
  detection, and analysis}.
\newblock Elsevier, 2013.

\bibitem{2}
``User monitoring with user behavior analytics.''
  \url{https://www.rapid7.com/solutions/user-monitoring/}.
\newblock Accessed: 2021-02-01.

\bibitem{3}
P.~L. Campbell, ``Department of defense instruction 8500.2 "information
  assurance (ia) implementation:" a retrospective.,''

\bibitem{bejtlich2013practice}
R.~Bejtlich, {\em The practice of network security monitoring: understanding
  incident detection and response}.
\newblock No Starch Press, 2013.

\bibitem{4}
Fortinet, ``Independent study pinpoints significant scada/ics security risks.''
  \url{https://www.fortinet.com/content/dam/fortinet/assets/white-papers/WP-Independent-Study-Pinpoints-Significant-Scada-ICS-Cybersecurity-Risks.pdf}.
\newblock Accessed: 2020-12-30.

\bibitem{9}
R.~Heenan and N.~Moradpoor, ``Introduction to security onion,'' in {\em The
  First Post Graduate Cyber Security Symposium}, 2016.

\bibitem{5}
Q.~Meng, D.~Li, and Y.~Ma, ``Research and application based on network security
  monitoring platform and device,'' in {\em 2019 IEEE Innovative Smart Grid
  Technologies-Asia (ISGT Asia)}, pp.~716--719, IEEE, 2019.

\bibitem{6}
S.~G. Mack and G.~Sriram, ``Netflow: A tool for isolating carbon flows in
  genome-scale metabolic networks,'' {\em Metabolic engineering
  communications}, vol.~12, p.~e00154, 2021.

\bibitem{7}
J.~Vacca, {\em Computer and information security handbook}.
\newblock Newnes, 2012.

\bibitem{8}
SecurityOnion, ``Securityonion, hardware requirements.''
  \url{https://docs.securityonion.net/en/2.3/hardware.html}.
\newblock Accessed: 2021-01-25.

\bibitem{elasticsearch2018elasticsearch}
B.~Elasticsearch, ``Elasticsearch,'' {\em Internet: https://www. elastic.
  co/pt/,[Sep. 12, 2019]}, 2018.

\bibitem{10}
C.~NETFLOW, ``Introduction to cisco ios netflow-a technical overview. cisco
  system,'' {\em Inc., http://goo. gl/BaQhxu}, 2007.

\bibitem{11}
S.~Jeong, J.-H. You, and J.~W.-K. Hong, ``Design and implementation of virtual
  tap for sdn-based openstack networking,'' in {\em 2019 IFIP/IEEE Symposium on
  Integrated Network and Service Management (IM)}, pp.~233--241, IEEE, 2019.

\bibitem{12}
L.-M. Wang, T.~Miskell, P.~Fu, C.~Liang, and E.~Verplanke, ``Ovs-dpdk port
  mirroring via nic offloading,'' in {\em NOMS 2020-2020 IEEE/IFIP Network
  Operations and Management Symposium}, pp.~1--2, IEEE, 2020.

\bibitem{14}
K.~Salah, M.~Hammoud, and S.~Zeadally, ``Teaching cybersecurity using the
  cloud,'' {\em IEEE Transactions on Learning Technologies}, vol.~8, no.~4,
  pp.~383--392, 2015.

\bibitem{15}
S.~Selim, A.~Amin, and Z.~Saad, ``Towards better computer game ai techniques,''
  in {\em 2nd Conference on Computer Science innovation technology (CCSIT)},
  2013.

\bibitem{16}
L.~Dawel, A.-K. Seifert, M.~Muma, and A.~M. Zoubir, ``A robust genetic
  algorithm for feature selection and parameter optimization in radar-based
  gait analysis,'' in {\em 2019 IEEE 8th International Workshop on
  Computational Advances in Multi-Sensor Adaptive Processing (CAMSAP)},
  pp.~674--678, IEEE, 2019.

\end{thebibliography}
\end{document}